\documentstyle[12pt]{article}\textwidth 17cm
\begin{document}
\large

\vspace{40mm}

\begin{center}
{\bf INFLUENCE OF GEOMAGNETIC PERTURBATION }\\ {\bf ON RESONANT
GRAVITATIONAL WAVE DETECTOR}\\

\vspace{5 mm}

{\bf P. Vorob'ev}\footnote{E-mail: vorobyov@inp.nsk.su}

{\small{\it Budker Institute of Nuclear Physics, Novosibirsk,
Russia 630090}}

{\bf V. Ianovski \footnote{E-mail: ianovski@lnpi.spb.su}}
and {\bf I. Okunev \footnote{E-mail: okunev@lnpi.spb.su}}

{\small {\it Petersburg Nuclear Physics Institute, Gatchina,
Russia 188350}}
\end{center}

\begin{quote}
{\small The level of background signals in modern  cryogenic  resonant
mass gravitational  wave  antenna  is  discussed  caused  by  (a)  the
geomagnetic field  pulsations  and  (b)  an  atmosferic  of  very  low
frequency band, generated by a lightning flash. The  analysis  of  our
results show that the signals of this origin will generally exceed the
signals  from  the  gravitational  wave  sources.  To  suppress  these
artifacts in such gravitational antenna, it is necessary  to  use  the
magnetometer included as anti-coincidence protection and a  system  of
magnetic screens.} \end{quote}

\vspace*{0.1cm} {\small PACS number(s): 04.80.N, 95.55.Y}

\section{Introduction}

In the last years the problem of detection of gravitational  waves  of
cosmic objects has moved to a qualitatively new level. The sensitivity
of such detector as resonant-mass  antenna,  working  at  temperatures
below $100~ mK$ should reach for short  bursts  the  range  of  $h\sim
10^{-21}$ \cite{1,2} and will come, in nearest future,  close  to  the
sensitivity of laser interferometer gravitational wave detectors, such
as in LIGO/VIRGO and LISA  experiments  \cite{3}.  The  transition  of
resonant-mass detectors to ultra low temperatures  operation  requires
to investigate the influence of low-frequency electromagnetic noise on
the gravitational wave antenna of this type.

The problem of electromagnetic background was discussed several  years
ago in an association with the results of  J.Weber  \cite{4}.  It  was
considered,  that  the  background  signals,  registered  by   Weber's
antenna, could be caused by an electromagnetic  noise  of  geomagnetic
origin. However Joseph Weber has experimentally demonstrated tolerance
of  his  gravitational  antenna  to  the  electromagnetic  signals  of
appropriate level as well as to  high  energy  cosmic  ray  particles.
Therefore this theme has not received further development.

Obviously, the influence  of  the  noise  of  such  origin  on  modern
resonant-mass detectors like ''NAUTILUS''  \cite{2}  is  qualitatively
different due to, first, essential increase of the sensitivity of  the
third  generation  detectors,  and,    secondly,    we    deal    with
superconductivity of the material in  these  generation  detectors  at
working  temperature  below  $1~K$.  The  veto  systems   there    are
practically in all gravitational detectors of  the  Weber's  type  for
elimination of a cosmic ray background. However detailed  analysis  of
narrow-band (bandwidth $~\sim 1~Hz$) resonant  antenna  excitation  by
geomagnetic field pulsations and lightning flash was not performed and
it is the main aim of this work.

\section{Influence of the magnetic field pulsation}

There are two mechanisms, linking the tension in a  solid  body  of  a
gravitational antenna with the external magnetic field:

1. For frequencies around $1~kHz$ the skin thickness in  aluminum  bar
is few mm, therefore the change of external magnetic field  with  this
(sufficiently high  frequency)  produces  the  appropriate  change  of
pressure, that can excite the gravitational antenna.

2. During the transition process in  the  superconducting  state,  the
multiton Al bar (the resonant bar is a cylinder, made usually  of  the
special alloy - Al 5056), can trap a significant part  of  a  magnetic
field flux of the Earth. For pure Al the transition temperature in the
superconducting state is $ T_c=1.18~K$,  and  Al  5056  alloy  becomes
superconducting at $T_c=0.925~K$ \cite{5}. Thus the external  magnetic
field pulsations will cause, by interaction with induced dipole moment
of the bar, an appreciable tension in the antenna body.

We shall evaluate both effects in this paper. It goes  without  saying
that we have to take into account the external  field  suppression  by
the vacuum tank and by the system of conducting cryogenic  shields  of
the gravitational detector.

We need to know the perturbation spectrum of the Earth magnetic  field
in the band around $1000~Hz$. The  magnetic  field  perturbations  are
determined in this band by the high-frequency tail of the  geomagnetic
field perturbation spectrum, caused, basically, by the interaction  of
the solar wind with the ionosphere, and by the low-frequency  tail  of
the atmospheric spectrum. Here the term  ''atmospheric''  denotes  the
transient field (electric or magnetic),  generated  by  the  lightning
flashes, or by any subsidiary features of the  flashes.  The  spectral
density of the geomagnetic field variations $S_H$  around  $1~kHz$  is
about $1\times 10^{-6}~A\cdot m^{-1}\cdot Hz^{-1/2}$ \cite {6}. It  is
useful to recall, that at the Earth average latitudes the spectrum  of
the magneto-telluric field of the atmospheric  has  a  maximum  around
$30~Hz$ \cite{6,7}.

The evaluation of spectral density of the magnetic field perturbations
is rather difficult in examined band at large distances,  because  the
low-frequency  tail  contains  peaks,  associated  with  the  resonant
properties of a cavity "earth-ionosphere" -  the  Schumann  resonances
\cite{8}. W. Schumann has predicted that the cavity formed between the
earth and the lower ionosphere should resonate at certain frequencies,
and that these resonances would be excited by the  lightning  flashes.
The  detailed  theory  of  Schumann  resonances  is  not  simple,  but
observations gave for the first  order  of  the  resonance  the  value
$7\div 9\;Hz$.

If we divide the frequency band from $5~Hz$ up to $2~kHz$  in  to  two
sites, then for the first of them $5\div 100\;Hz$ we can write \cite{6}
\begin{equation}
\label{1}S_H(f)\simeq 4\times 10^{-7}\cdot f^{0.3}~,
\end{equation}
and for second, $100\div 2000\;Hz$
\begin{equation}
\label{2}S_H(f)\simeq \frac 1f\cdot 10^{-3}~.
\end{equation}

More detailed theory  of  geomagnetism,  including  the  data  on  the
magnetic field perturbations, can be found  in  \cite{6,9,10,11}.  The
top of the geomagnetic field perturbation  spectrum  -  the  frequency
band above $1.5-2~kHz $  is  connected,  basically,  with  ionospheric
processes and with the influence of the space radiation. The  spectral
density on these frequencies are,  with  95  \%  probability  ,  below
$1\times  10^{-8}~A\cdot  m^{-1}\cdot  Hz^{-1/2}~$  \cite{6,11}.    It
follows  from  estimations  Eq.(\ref{1},\ref{2}),  that  the  spectral
density  of  the  magneto-telluric  field  does  not  exceed  $1\times
10^{-6}$ in all the interesting frequency band.

It  is  necessary  to  add  some  words  about  the  magnetic    field
perturbation by the near lightning  flash.  Note,  that  the  spectral
density of magnetic component of the electromagnetic wave in ULF  band
strongly depends on the lightning flash distance and  its  orientation
\cite{12}.  The  current  in  lightning   reaches    $1\div    2\times
10^5\;A$\cite{12,13}, at distances more than $10~km$ it is not  already
a linear conductor with a current, which we use as the model only  for
valuation (as the field will fall appreciably faster).  There  is  the
well-known expression for the current distribution along  the  channel
in double-exponential form \cite{12}
\begin{equation}
\label{3}i_t=i_0[\exp (-at)-\exp (-bt)].
\end{equation}
Here $a=2\times 10^4~s^{-1},b=2\times 10^6\ s^{-1}$.

For  an  incoming signal the spectral density $S(f)$ is defined by
\begin{equation}
\label{4}S(f)\sim \frac{(a-b)}{\sqrt{(a^2+f^2)(b^2+f^2)}}~.
\end{equation}

We shall evaluate the value of the horizontal magnetic flux density of
the thunderstorm $H_{ts}\ $ now. In these  circumstances  $H_{ts}$  is
given approximately by \cite{13}
\begin{equation}
\label{5}H_{ts}\approx \frac 1{4\pi R^2} \frac{dM}{dt}+\frac1
{4\pi cR} \frac{d^2M}{dt^2}~,
\end{equation}
where $R$ is the distance from the source, $c$ - the speed  of  light.
The "charge" moment at any time $t$  is  given  by  $M_t=2\sum  q_zz$,
where the summation covers all elementary charge of magnitude $q_z$ at
height $z$.

If  $R=10~km$,  and  $I=1\times  10^5~A$,  we  can  obtain  now  using
Eq.~(\ref{3}, \ref{4}, \ref{5})
\begin{equation}
\label{6}H_{ts}\simeq \frac I{2\pi \cdot R}\simeq 1~,
\end{equation}
and the spectral density will be:
\begin{equation}
\label{7}S_{ts}(f)\simeq 1\times 10^{-3}\div 1\times 10^{-4}.
\end{equation}
It  can  be  seen  that  the  influence  of  the   thunderstorms    is
approximately 3---4 orders of magnitude higher  than  the  geomagnetic
field variations.

Let the geomagnetic field H$_{geo}$, acting on a cylinder, consist  of
two terms: the constant component $H_0$ equal to the average intensity
of the Earth field in a given place and $H(\omega )$, the variable part

\begin{equation}
\label{8}H_{geo}=H_0+H(\omega)
\end{equation}
Here the value of the Earth magnetic field is  $H_0\simeq  40~A  \cdot
m^{-1}$. For simplicity we shall consider the vector $H_{geo}$  to  be
perpendicular to the cross section plane of  the  cylindric  bar  with
area $S$, (however this is not an  essential  restriction).  Then  the
force $F$ of the geomagnetic field pressure on the detector end face is

\begin{equation}
\label{9}F  =  \mu_0H_{geo}^2 \cdot S = \mu_0 \left[
H_0^2+2H_0H(\omega)+ H^2(  \omega) \right] \cdot S,
\end{equation}
where $\mu_0=4\pi\times 10^{-7}~H \cdot m^{-1}$ is the permeability of
the empty space.

The first term in Eq.(\ref{9}) corresponds to the static  pressure  of
the magnetic field, which rises when we consider  the  superconducting
cylinders only. In our case it is not of interest, since  it  deviates
the resonant frequency very slightly. It is possible to  neglect  also
the third term here, the electro-magnetic wave pressure,  due  to  its
very small value. As a result we receive
\begin{equation}
\label{10}F(\omega )=\mu _0H_0\cdot H(\omega )\cdot S~.
\end{equation}

The force $F$ describes here the action of  the  variable  geomagnetic
field on any detector or shield of a well  conducting  material,  when
the size of the antenna bar is much more than the skin-layer thickness
on the resonant frequency. For resonant gravitational antenna from the
superconducting material the force of the electromagnetic pressure  at
resonant  frequency  is  determined  by  the  same   expression,    as
Eq.(\ref{10}). It does not depend, whether the  Earth  magnetic  field
flux was trapped  by  the  superconducting  antenna  bar,  or  it  was
completely superseded at cooling.

On the main frequency mode of the  ultracryogenic  resonance  detector
NAUTILUS \cite{2} (about 1kHz), the geomagnetic noise is rather small:
$S_H\mid_{f=1kHz}<10^{-6}$. The cross section area of cylindrical  bar
$S\simeq 1~m^2$. Therefore, the force of the electromagnetic  pressure
$F_{em}$ (in $1~Hz$ bandwidth ) by the action of geomagnetic noise can
be roughly described as
\begin{equation}
\label{11}F_{em}=\mu _0H_0\cdot H\mid _{1kHz} \cdot S \simeq
1\times 10^{-11}~N.
\end{equation}
And the magnetic perturbations by lightning flashes produces the force
$F_{ts}(\omega)$
\begin{equation}
\label{12}F_{ts}(w)=\mu _0H_{ts}(\omega )\cdot H_0 \cdot S\approx
1\times 10^{-7}\div 1\times 10^{-8}~N.
\end{equation}

Let us compare the perturbations  excited  by  the  geomagnetic  field
variation with  the  signal  from  the  gravitational  wave.  Just  to
simplify  this  task  let  the  mass  of  the  cylindrical  bar    be:
$m=1000~kg$,  length  $L=1~m$,  cross-section   $S=1~m^2$,    resonant
frequency $f=1000~Hz$ and the bandwidth $\Delta f_d=1~Hz$. And just to
set the scale of estimates set  the  detector  sensitivity  $h=1\times
10^{-20}$. We shall evaluate the excitation of the conducting  Al  bar
by  geomagnetic  field  perturbations,    leaving    out    the    bar
superconductivity (at zero freezing flux).

The value of $F_g$, acting on the bar body, is:
\begin{equation}
\label{13}F_g(f_g,\Delta f_d)\simeq  \pi  \cdot  mL\int  f^2\cdot
Z\cdot df_d~,
\end{equation}
where $f_g$ --- the frequency of gravitational wave,
$\Delta f_d$ --- the bandwidth of gravitational antenna and
$Z$ --- the  Fourier  image of h.

For the gravitational burst of sine shape with duration of the  packet
of waves $~\tau_g$ we have
\begin{equation}
\label{14}Z=h\cdot \frac{\tau _g}2\left\{ \frac{\sin (f-f_g)\cdot \tau
_g/2}{(f-f_g)\tau _g/2}\right\} ~.
\end{equation}
After integration Eq.(\ref{13}) over the bandwidth  of  the  resonance
detector with resonance frequency $f_d$ we have
\begin{equation}
\label{15}F_g\simeq \pi \cdot mLh\cdot f_d^2\cdot \tau _g\cdot \Delta
f_d\left\{ \frac{\sin (f_d-f_g)\cdot \tau _g/2}{(f_d-f_g)\tau _g/2
}\right\}~,
\end{equation}
that in the presence of small frequency deviation $(f_d-
f_g)\cdot \tau_g<<1$ makes
\begin{equation}
\label{16}F_g\simeq \pi \cdot mLh\cdot f_d^2\cdot \tau_g \cdot \Delta f_d
\end{equation}
For short pulses of gravitation radiation, when $\tau_g\sim 1/f_d\sim
1~ms$,we have
\begin{equation}
\label{17}F_g\simeq 10^{-13}~N.
\end{equation}

Let us equate this value to the magnetic field pressure $F_{mf}$, that
is necessary for reception of a comparable signal. We have
\begin{equation}
\label{18}F_{gr}\simeq F_{em}=\mu_0H_0\cdot H(\omega )\cdot S.
\end{equation}
Using this equation, we can find the corresponding estimate for $H$
\begin{equation}
\label{19}H\simeq 1\times 10^{-7}.
\end{equation}

Comparison of Eq.(\ref{11},\ref{12}) and Eq.(\ref{17}) shows that  the
stress  in  the  detector  body  due  to  its  interaction  with   the
geomagnetic field perturbations (and more, with the fields of  near-by
thunderstorm)  will  exceed  the  stress  due  to  the   gravitational
wave-induced tide! Certainly, it is necessary to  shield  the  antenna
bar and to arrange a veto-system connected with the magnetometer.

\section{Shielding}

We shall estimate the shielding property of the vacuum tank and of the
system of heat shields, surrounding the detector bar. The  designs  of
the modern  third  generation  ultracryogenic  resonant  antennae  [2]
require the vacuum tank and the cryostat walls of stainless steel, the
cryogenic shields of copper.

Let the thickness of the skin-layer at the frequency $\omega $ be  $d$
and we get
\begin{equation}
\label{20}d=\frac c{\sqrt{2\pi \cdot \sigma \cdot \mu \cdot \omega }}\;.
\end{equation}
Then for the copper (at frequency 1 kHz) the skin thickness  is  about
1.5 mm, and for stainless steel becomes about 5 mm. In the case of the
thickness of the  wall  $h>d$,  the  attenuation  coefficient  of  the
magnetic field $k_1$ is defined by \cite{15,16}

\begin{equation}
\label{21}k_1=C\cdot \left( \frac S{Ld}\right) \cdot e^{h/d}.
\end{equation}
here $C$ is a coefficient about 1, depending on the tank geometry; $S$
- the cross-section orthogonal to magnetic field; $L$ - the  perimeter
of this cross-section. In the case of $h<d$ we  have  the  attenuation
coefficient $k_2$ as
\begin{equation}
\label{22}k_2=\sqrt{1+\left( \frac{2Sh}{Ld^2}\right)^2}.
\end{equation}

It is interesting to note, that  for  the  vacuum  tank  of  stainless
steel, and also for the copper heat shields h and d are the values  of
the same order of magnitude. For such case we shall make estimates for
both, Eq.(\ref{21} and  \ref{22}),  at  $h=d$.  Let  $S=2~m^2,~L=6~m,\
d=0.005~m$, then we have
\begin{center}
\begin{equation}
\label{23}
\begin{array}{c}
k_1\simeq 300, \\
k_2\simeq 240.
\end{array}
\end{equation}
\end{center}

As it is seen from Eq.(\ref{23}), these  estimates  are  well  agreed.
Thus, the attenuation  coefficient  of  the  magnetic  field  for  the
detector  vacuum  tank  is  $200\div  300$  on  the  frequency   about
$1000\;Hz$. However, the vacuum tank, also the thermal shields are, as
usual, not continuous  and  contain  flanges  and  other  elements  of
design, hindering the induced currents. As the experiments  show,  the
shielding factor (in the direction of field )  appreciably  drops  for
flanges with dimensions of the same order as the  tank  cross-section.
The experimental values of the shielding factor k for flange with  the
rubber  sealing  joints  are  $k=30\div  300$,  depending  on  design.
Meantime for  the  metal  sealing  joints  we  have  $k>300$  and  the
demountable  flange  joint  does  not,  practically,  deteriorate  the
shielding  properties  of  a  vacuum  tank.  In  the  same  time,  the
longitudinal cuts (along the cylinder copper heat shields), can reduce
the shielding factor  of  a  separate  shield  up  to  10  times.  The
shielding factor of the multilayer shield is always  lower,  than  the
product of shielding factors of separate layers and is  determined  by
its mutual arrangement and by the  shield  dimensions.  Therefore  the
definition of the field reduction factor requires complex calculations
or experimental measurements. However for  multilayer  heat  shielding
systems $~k$ is, to all appearance, always more than 100.

We have conducted also the actual measurements for 3 helium  cryostats
of slightly different types with the typical dimensions: length  about
1.5 m, external diameter 0.5 m, internal diameter 0.3 m.  Walls  were:
$aluminum$ $alloy$ $3 mm$ $+ copper$ $2 mm$ $+ stainless$ $steel$ $1.5 mm$.
The stainless steel flange covers the top end of the cryostat. For our
measurements the cryostat  was  placed  in  a  special  solenoid.  The
magnetic field was excited at the frequency 1000 Hz and  was  measured
inside and outside of the cryostats by an inductive probe,  which  was
sensitive only to the magnetic field. The probe signal was measured by
the spectrum analyzer SK4-56, which was used as  the  high-sensitivity
narrow-band selective voltmeter. It is clear, that the  ratio  of  the
magnetic field outside to the field inside the cryostat gives  us  the
shielding factor. The shielding factor of the magnetic  field  was  in
the range between 300 and 500 for all the  studied  cryostats  in  our
experiments.

\section{Conclusion}

Our analysis result  show,  that  the  tensions,  connected  with  the
geomagnetic field perturbation will  surpass  the  gravitational  wave
influence. A good system of magnetic screens, that are able to ease an
external magnetic field at frequencies near to the  working  frequency
of the detector not less than three orders, is necessary in that  case
as minimum. Certainly, the best way to do it --- to take into  account
at  the  detector  design.  It  is  necessary,  practically,    either
compensate or shield the Earth field (by compensating coils or by  the
magnetic shields). The most  natural  way  to  suppress  the  magnetic
background is to arrange the complete installation into  the  magnetic
shield, so-called magnetic room. The last way is more reliable, but it
will be more expensive. Nevertheless, the 2-3 layer magnetic shield
permits to suppress the earth magnetic field (and all its fluctuations
too) not less  than  3  orders  of  magnitude  and  more  without  any
monitoring systems. However, note  once  more, that  the  modern
gravitational  detectors
operate typically in the helium temperature range - below 1.8 K.  Thus
such detector needs a special cooling system  with  heat  bridges  and
with broad cuts in the cryogenic shields and in the  tank  walls.  All
these elements of the installation construction can to deteriorate the
magnetic shielding and to pass an electromagnetic damage, which
can be the sources of the noise of nonstationary nature, additional to
the fundamental sources of the noise in the detector \cite{1}. In  this
context to suppress artifacts  from  close  lightning  flashes  it  is
necessary,according to our results,  to  use  a  magnetometer  system,
which should be included as an anti-coincidence protection  from  false
signals of terrestrial origin. In particular it is important  for  the
near future gravitational measurements  in  coincidence  with  several
ultralow temperature antennas - the AURIGA  antenna,  INFN  Laboratori
Nazionali di Legnaro near Padova, the antenna in  preparation  at  the
Stanford University and the NAUTILUS antenna of the Rome group.

To all this it should be added, that the antenna bar  is  making  some
noise during the cooling process due to relieving of the thermo-stress.
The complete adiabatic change  of  magneto-telluric  field  can  become  a
trigger mechanism that will provoke a false signal. The cause of such
shakes is the removal of stresses, for which  there  was  a  potential
barrier at usual  temperature, and  which  were  "captured" in local
minima of potential stress energy.

Apparently, one more source of noise exists. It arises because of the
strong magnetic coupling of the heat screen to the antenna bar,
that could excite variable tensions in the antenna bar due to
oscillations in the heat screen, caused by external perturbations,
such as the acoustic vibration, electromagnetic field, etc. Thus all
these will  arise induced vibrations of the installation. The noise
of  such  origin will be present, most probably, also in  the
gravitational wave detectors like VIRGO/LIGO \cite{2,3}.

We will continue the study of both these sources of terrestrial origin
noise in the gravitational wave detector.

\vspace*{0.2cm}

We would like to express our  gratitude  to  all  the  colleagues  and
friends who  helped  us  in  different  ways  during  this  study.  In
particular we thank Eugenio Coccia (Roma), Igor  Kolokolov (Novosibirsk),
Valentin Rudenko  (Moscow).  We  also  thank our friends Stanly Forgang
and Gregory  Itskowitch  (Texas)  for  very efficient discussions.
One of us (P.V.) thanks the "Cosmion" Center  for financial support.

\end{document}